\documentclass[a4paper,11pt]{article}
\pdfoutput=1 

\usepackage{jheppub} 

\usepackage[T1]{fontenc} 
 \usepackage{graphicx}
\usepackage{amsmath}
\usepackage{amssymb}
\usepackage{array}
\usepackage{slashed}
\usepackage{comment}
\usepackage{datetime}
\usepackage{soul}
\usepackage{pifont}

\usepackage{float}
\usepackage{bm}
\usepackage{bbm,multicol}

\usepackage{subfigure}
\usepackage{dsfont}

\usepackage{marginnote}

\usepackage{multirow}

\def\h{h^0}
\def\H{H^0}
\def\A{A}
\newcommand{\sba}{\ensuremath{\sin(\beta-\alpha)}}
\newcommand{\cba}{\ensuremath{\cos(\beta-\alpha)}}
\newcommand{\hc}{H^{\pm}}
\newcommand{\w}{W^{\pm}}

\newcommand{\ifb}{\ensuremath{ \text{fb}^{-1}  }}
\newcommand{\cmark}{\ding{51}}%
\newcommand{\xmark}{\ding{55}}%

\newcommand{\minigraph}[5][0.25in]{\begin{minipage}{#2}\begin{center}\includegraphics[width=#2]{#5}\\\vspace{#3}\hspace{#1}{\footnotesize #4}\end{center}\end{minipage}}

\title{Charged Higgs Search via $AW^\pm/HW^\pm$ Channel}
\author{Baradhwaj Coleppa, Felix Kling, Shufang Su}
\emailAdd{baradhwa@email.arizona.edu, kling@email.arizona.edu, shufang@email.arizona.edu}
\affiliation{Department of Physics,
University of Arizona,
Tucson, AZ 85721, USA}

\abstract{
Models of electroweak symmetry breaking with extended Higgs sectors are theoretically well motivated. In this study, we focus on models with a low energy spectrum containing a pair of charged scalars $\hc$, as well as a light scalar $H$ and/or a pseudoscalar $A$.  We study the $\hc tb $ associated production with  $\hc \rightarrow A\w/H\w$, which could reach sizable branching fractions in certain parameter regions. With detailed collider analysis, we obtain the exclusion bounds as well as discovery reach at the 14 TeV LHC for the process $p p \to  \hc  tb \to  A\w tb/ H\w tb \rightarrow \tau\tau bbWW, bbbbWW$.   We find that for a daughter particle mass of  70 GeV, the 95\% C.L. exclusion reach in $\sigma\times{\rm BR}$  varies from  about  60 fb to 25 fb, for $m_{\hc}$ ranging from 150 GeV to 500 GeV with 300 ${\rm fb}^{-1}$ integrated luminosity in the $\tau\tau$ mode. We further interpret these bounds  in the context of  Type II Two Higgs Doublet Model.     The  exclusion region in the $m_{\hc}-\tan\beta$ plane can be extended to $m_{\hc}=$ 600 GeV, while discovery is possible for $m_{\hc}\lesssim$  400 GeV with 300 ${\rm fb}^{-1}$ integrated luminosity.    The  exotic decay mode   $\hc \rightarrow A\w/H\w$ offers a complementary channel to the conventional mode   $\hc \rightarrow \tau\nu$  for   charged Higgs searches.   
}

\begin{document}

\maketitle
\flushbottom
\newpage

\section{Introduction}
\label{sec:intro}

The discovery of the Standard Model (SM)-like Higgs at the LHC \cite{Aad:2012tfa, ATLAS:2013sla, Chatrchyan:2012ufa,CMS:yva} marks the final and one of the most important discoveries within the SM  of particle physics as regards its particle content. The ATLAS and CMS experiments have reported precise measurements of the mass of this particle, as well as the determination of  its spin \cite{ATLAS:2013sla,CMS:yva,Aad:2013xqa}. The present scenario raises interesting questions about the origin of Electroweak Symmetry Breaking (EWSB). It is conceivable that the scalar sector of the SM does indeed engineer all of EWSB, but at the same time we have compelling evidence from theoretical and experimental fronts that the SM needs to be supplanted with other dynamics for it to consistently explain issues like the naturalness problem,  neutrino masses and the dark matter in the universe. Thus it is entirely possible that the scalar sector of the SM responsible for EWSB itself has a richer structure. Early attempts toward  enlarging the  scalar sector resulted in the Two Higgs Doublet Models (2HDM)~\cite{Branco:2011iw,type1,hallwise,type2}. Other examples also involving an enlarged scalar sector include the Minimal Supersymmetric Standard Model (MSSM) \cite{Nilles:1983ge,Haber:1984rc,Barbieri:1987xf} and the Next to Minimal Supersymmetric Standard Model (NMSSM) \cite{Ellis:1988er,Drees:1988fc}.

Models with extended Higgs sectors hold a lot of phenomenological interest. The discovery of extra Higgs bosons  would serve as unambiguous evidence for new physics beyond the SM. A clear indication for a non-minimal Higgs sector as a source of EWSB would be the observation of charged Higgs bosons $\hc$ which are absent in the SM.   The discovery of the charged Higgs, however, is quite challenging at colliders.  If the mass of the charged Higgs $m_{\hc}$ is smaller than the top mass $m_t$, the dominant production mechanism of the charged Higgs is via top decay: $t \rightarrow bH^{+} $. Most studies performed at LEP, Tevatron and LHC focus on such light charged Higgs bosons which are assumed to either decay leptonically ($\hc \rightarrow \tau \nu$),  or into jets ($\hc \rightarrow c s$).    In the case of a heavy charged Higgs with $m_{\hc} > m_{t}$, the main production mode is the top quark associated production $\hc tb$.    For the dominant decay $\hc \rightarrow tb$, it is difficult to identify the $ttbb$  signal  given the huge irreducible SM backgrounds.  The current heavy charged Higgs searches thus mostly focus on the subdominant decays $\hc \rightarrow \tau\nu$ or $cs$  in order to take advantage of the cleaner signal and suppressed backgrounds.

Other possible decay channels like $\hc \rightarrow AW^\pm, HW^\pm$ open up once they are kinematically accessible,  where  $H$ and $A$ refer to the generic CP-even and CP-odd Higgs,  respectively\footnote{Note that we use $\h$ and $\H$ to refer to the lighter or the heavier CP-even Higgs for models with two CP-even Higgs bosons.  When there is no need to specify, we use $H$ to refer to the CP-even Higgses.}.  In the 2HDM, the couplings $\hc A W^\mp / \hc H W^{\mp}$ are controlled by the electroweak gauge coupling $g$.  While the coupling to $A$ is independent of the mixing angles, the coupling to $H$ is maximized for non-SM-like CP-even Higgses.  These exotic decays quickly dominate over $\tau\nu, cs$ once they are open, and could be even larger than the $tb$ mode for a large range of $\tan\beta$.    It was shown that in the 2HDM or NMSSM, both decays $\hc \rightarrow A_iW^\pm, H_iW^\pm$ could appear with large branching fractions\footnote{$\hc \rightarrow AW^\pm,  HW^\pm$ is less likely to open in the  MSSM due to kinematical constraints that force $m_{\hc} \sim m_A \sim m_{H}$ for the non SM-like Higgses.}~\cite{Chiang:2013ixa,MODEL_NMSSM, Grinstein:2013npa}. It is thus timely to study such charged Higgs decay channels and fully explore the experimental discovery potential for an enlarged Higgs sector. 

In this paper, we focus on $\hc tb $ associated production of the charged Higgs with the subsequent exotic decay of  $\hc \rightarrow A\w/H \w$. We consider leptonic decay of one of the $\w$ either coming from $\hc$ or top decay, with the $A/H$ in the final state decaying into a pair of fermions ($bb$ or $\tau\tau$) and explore the exclusion bounds as well as the discovery reach at the LHC for various combinations of $(m_{\hc},m_{H/A})$. ATLAS investigated this decay mode in an early study \cite{ATLAS_EARLY_STUDY, Assamagan:2002ne} focusing on the $A/H \to bb$ mode only. So far no analysis has been done for the more promising $A/H\to\tau\tau$ mode. 

 A light charged Higgs could have a large impact on precision and flavor observables~\cite{FLAVOR}. For example, in  the 2HDM, the bounds on $b\to s\gamma$ restrict the charged Higgs to be heavier  than 300 GeV. A detailed analysis of precision and flavor bounds in the 2HDM can be found in Refs.~\cite{Coleppa:2013dya,Mahmoudi:2009zx}. Flavor constraints on the Higgs sector are, however, typically   model-dependent, and could be alleviated when there are contributions from other new particles in the model. Our focus in this work is   on collider searches for the charged Higgs and its implications for the Type II 2HDM.  Therefore, we consider a wide range of  charged Higgs mass.

The paper is organized as follows.  In Sec.~\ref{sec:scenarios}, we present a brief overview of models and parameter regions  where $\hc \rightarrow A\w/H \w$  can be significant.  In Sec.~\ref{sec:limits}, we summarize the current experimental search limits on charged Higgses. Sec.~\ref{sec:analysis} describes the collider analysis in detail. After describing the signal process and event generation in Sec.~\ref{sec:signal},  we present the details of the analysis for the $A/H \to \tau\tau$ channel in Sec.~\ref{sec:tautau}.  We   show the model independent results of 95\% C.L. exclusion as well as 5$\sigma$ discovery limits for $\sigma(pp \rightarrow  \hc tb  \rightarrow  A/H \w tb \to \tau\tau bbWW)$ at the 14 TeV LHC with 100, 300 and 1000 ${\rm fb}^{-1}$ integrated luminosity.  In Sec.~\ref{sec:ana_bb} we present the analysis for the  $H/A \to bb$ final state and derive the corresponding cross section limits.   In Sec.~\ref{sec:implication}, we study the implications of the collider search 
limits on the  Type II 2HDM. We conclude in Sec.~\ref{sec:conclusions}.

 \section{Scenarios with large $H^\pm \rightarrow A\w /H\w$ }
 \label{sec:scenarios}

In the 2HDM,  we introduce two ${ \rm SU}(2)_L$ doublets   $\Phi_{i}$,  $i=1,2$:
 \begin{equation}
\Phi_{i}=\begin{pmatrix} 
  \phi_i^{+}    \\ 
  (v_i+\phi^{0}_i+iG_i)/\sqrt{2}  
\end{pmatrix},
\label{eq:doublet}
\end{equation} 
where $v_1$ and $v_2$ are the vacuum expectation values (vev) of the neutral components which satisfy the relation $\sqrt{v_1^2+v_2^2}=$ 246 GeV after EWSB.  Assuming an additional discrete ${\cal Z}_2$ symmetry imposed on the Lagrangian,  we are left with six free parameters, which can be chosen as  the four Higgs masses ($m_{\h}$, $m_{\H}$, $m_A$, $m_{H^{\pm}}$), a mixing angle $\alpha$ between the two CP-even Higgses, and the ratio of the two vacuum expectation values, $\tan\beta=v_2/v_1$.   In the case where a soft breaking of the ${\cal Z}_2$ symmetry is allowed, there is an additional parameter $m_{12}^2$.  

The Higgs mass eigenstates containing a pair of CP-even Higgses $(\h, \H)$, one CP-odd Higgs $\A$ and  a pair of charged Higgses $H^\pm$   can be written as\footnote{For more details about the 2HDM  model, see Ref.~\cite{Branco:2011iw}.}:
\begin{equation}
\left(\begin{array}{c}
\H\\ \h
\end{array}
\right)
=\left(
\begin{array}{cc}
\cos\alpha &\sin\alpha\\
-\sin\alpha&\cos\alpha
\end{array}
\right)  \left(
\begin{array}{c}
\phi_1^0\\\phi_2^0
\end{array}
\right),\ \ \ 
\begin{array}{c}
 A \\H^\pm
 \end{array}
 \begin{array}{l}
 =  -G_1\sin\beta+G_2\cos\beta\\
 =-\phi_1^{\pm}\sin\beta+\phi_2^{\pm} \cos\beta
 \end{array}.
 \label{eq:mass}
 \end{equation}
The couplings that are of particular interest are of the type $\hc A W^{\mp}$ and $\hc H W^{\mp}$. They are determined by the gauge coupling structure, as well as the mixing angles \cite{Gunion:1989we}:   
\begin{eqnarray}
 g_{\hc h^0 W^{\mp}}&=& \frac{g\cba}{2}(p_{\h}-p_{\hc})^\mu,  \\
 g_{\hc H^0 W^{\mp}}&=& \frac{g\sba}{2}(p_{\H}-p_{\hc})^\mu,  \\
 g_{\hc A W^{\mp}}&=& \frac{g}{2}(p_{A}-p_{\hc})^\mu,  
 \label{eqn:hc-couplings}
 \end{eqnarray}
 with $g$ being the ${\rm SU}(2)_L$ coupling,  and $p_\mu$ being the incoming momentum for the corresponding particle.

An interesting feature here is that $\hc$ always couples to the non-SM-like CP-even Higgs more strongly. If we demand $\h$ ($\H$) to be SM-like, then $|\sba|\sim 1$ ($|\cba|\sim 1$), and the $\hc \H W^{\mp}$ ($\hc \h W^{\mp}$) coupling is unsuppressed.  Therefore, in the $\h$-126 case,  $\hc$ is more likely to decay to $\H \w$ than $\h \w$  unless the former decay is kinematically suppressed.  In the $\H$-126 case,  $\hc$ is more likely to decay to $\h \w$ than $\H \w$.  The $\hc A W^{\mp}$ coupling, on the other hand,  does not depend on any mixing angle and therefore this decay is not suppressed once it is kinematically allowed. 

In the generic 2HDM, there are no mass relations between the charged scalars, the scalar and pseudoscalar states.  Thus, the decays $\hc \rightarrow \h \w,\ \H \w$ and $\hc \rightarrow A \w$ can all be kinematically accessible and  dominate in different regions of parameter spaces. It was shown in Ref.~\cite{Coleppa:2013dya} that in the Type II 2HDM with ${\cal Z}_2$ symmetry, imposing all experimental and theoretical constraints still left sizable regions in the parameter space that permit such exotic decays with unsuppressed decay branching fractions.   

The dominant competing mode is $\hc \to tb$, which is controlled by the $\hc t b$ coupling
\begin{equation}
g_{\hc tb} = \frac{g}{2 \sqrt{2} m_W} \left[ (m_b \tan \beta + m_t \cot \beta ) \pm (m_b \tan \beta - m_t \cot \beta ) \gamma_5 \right]
\label{eqn:htbvertex}
\end{equation}	
in the Type II 2HDM.  At both small and large $\tan\beta$, $\Gamma(\hc \to tb)$ is increased  given the enhanced top and bottom Yukawa coupling, respectively.    The subdominant channel   $\hc \to \tau\nu$ has similar enhancement at large $\tan\beta$ as well.  
 
\begin{figure}[h!]
\centering
	\includegraphics[scale=1.0]{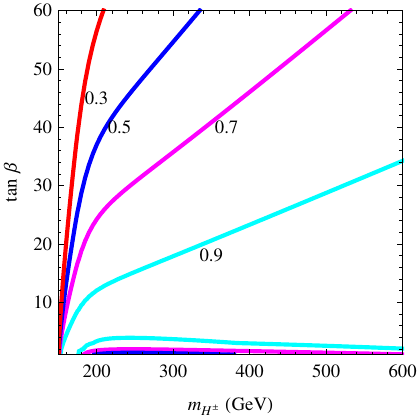}
	\includegraphics[scale=1.0]{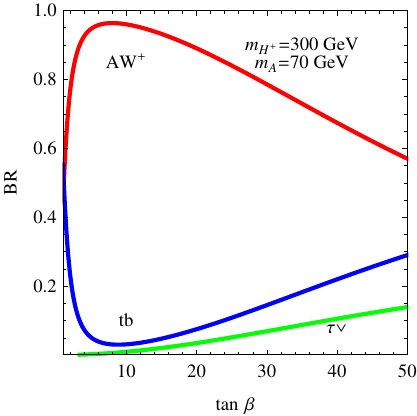}
\caption{Left panel: Branching fraction BR$(\hc\to A\w)$ in the $m_{\hc}-\tan\beta$ plane, for $m_A=$  70 GeV and   $\sba=$1.  Right panel: The branching fractions of $\hc$ as a function of $\tan\beta$ for   various decay modes: $\hc \rightarrow A \w$ (red), $tb$ (blue),  $\tau\nu$ (green) for $m_{\hc}=300$ GeV, $m_A= 70$ GeV and $\sba=1$.   }
\label{fig:HpBR}
\end{figure}
 
 In the left panel of Fig.~\ref{fig:HpBR}, we present contours of the branching fraction BR$(\hc\to A\w)$ in the $m_{\hc}-\tan\beta$ plane fixing $\sba=$ 1, $m_A=$  70 GeV and decoupling   $\H$.   It is seen that there is a ``kink'' at the $tb$ threshold which brings down the steeply increasing values of BR$(\hc\to\A\w)$.  Even so,  the $\A\w$ mode can be 90\% or higher in the band $1.5\lesssim\tan\beta\lesssim 30$ for  $m_{\hc}$ between 175  and 600 GeV. For large or small values of $\tan\beta$,  BR$(\hc\to A\w)$  is reduced due to   competition from  $\hc \rightarrow tb, \tau\nu$  modes. The $\hc\to \H\w$ mode, when kinematically accessible,  would show similar features with additional phase space suppression.  $\hc \rightarrow \h \w$ mode is maximized at $\sba=$ 0, which could be  a potentially useful search channel for $\hc$ in  the $\H-$ 126 case.     The current searches for the charged Higgs focus on the  $\hc \rightarrow \tau\nu$ channel, which is  sensitive to the   large $\tan\beta$ region.   We expect the $\hc\to A\w/H\w$ channel to be complementary for small or intermediate  $\tan\beta$. 
 
In   the right panel of Fig.~\ref{fig:HpBR}, we show the branching fractions of $\hc$ as a function of $\tan\beta$ for various decay modes of $\hc \rightarrow A \w,\ tb, \tau\nu$ for $m_{\hc}=300$ GeV, $m_A=$  70 GeV and $\sba=1$. For  almost all values of $\tan\beta$, the decay to the  $\A\w$ mode exceeds that of $tb$.   
 
The Higgs sector in the MSSM is more restricted, given that the quartic Higgs couplings are fixed by the gauge couplings and the tree-level Higgs mass matrix only depends on $m_A$ and $\tan\beta$. 
The decay $\hc \to \h \w$ is typically suppressed by the small coupling $\cba\sim 0$, and is only relevant for small $\tan\beta$.    The branching fraction is typically about 10\% or less \cite{Heinemeyer:2013tqa}.  In the usual decoupling region with large $m_A$,  the light CP-even Higgs $h^0$ is SM-like while the other Higgses are almost degenerate: $m_{\H} \sim m_{A} \sim m_{H^\pm}$.  Thus,  $\hc\to \H \w$ or $\hc \to A \w$ is not kinematically allowed. However, it has been shown that there are scenarios with large $\mu$ in which next-to-leading order (NLO) corrections can increase the mass difference between the charged and neutral Higgses~\cite{MODEL_HEAVY_H}, which could make this channel kinematically accessible.   In the NMSSM, the Higgs  sector of MSSM is enlarged to include an additional singlet. It was shown in Ref.~\cite{MODEL_NMSSM} that in this model, there are regions of parameter space where the decay $\hc \to H_i \w / A_i \w$ can be significant.

 \section{Current limits}
 \label{sec:limits}

Searches for a light charged Higgs boson with mass $m_{\hc}<m_t$ have been performed both by ATLAS and CMS~\cite{TheATLAScollaboration:2013wia,CMS_taunu} with 19.7 $\ifb$ integrated luminosity at 8 TeV and 4.6 $\ifb$ integrated luminosity at 7 TeV.  The production mechanism considered is top pair production in which one top quark decays into a charged Higgs $t \rightarrow b \hc$ while the other top decays into $bW$.  Assuming a branching fraction BR$(\hc \rightarrow \tau \nu) = 100 \%$,  the null search results from CMS~\cite{CMS_taunu} imply an upper bound for the top quark branching fraction BR$(t \rightarrow b \hc) = 1.2\%$  to 0.16\% for charged Higgs masses between 80 GeV and 160 GeV.  This result can be translated into bounds on the MSSM parameter space. In the $m_h^{\rm max}$ scenario of the MSSM, this excludes $ m_{\hc} <  155$ GeV for all values of $\tan\beta$.   Only the small region 155 GeV $<  m_{\hc} < 160 $ GeV around $\tan \beta = 8$  is still allowed.  The ATLAS results \cite{TheATLAScollaboration:2013wia} are similar.
 
A search with the $\hc \rightarrow cs$ final state has been performed by ATLAS \cite{Aad:2013hla} using 4.7 fb$^{-1}$ integrated luminosity at 7 TeV. Assuming BR$(\hc \rightarrow c s) = 100 \%$,  this implies an upper bound for the top quark branching fraction BR$(t \rightarrow b \hc ) = 5\%$ to  1\% for charged Higgs masses between 90 GeV and 150 GeV.   
 
Both ATLAS and CMS have also searched for a heavy charged Higgs boson with mass $m_{\hc}>m_t$ produced in association with a top quark~\cite{TheATLAScollaboration:2013wia,CMS_taunu}.  With 19.5 $\ifb$ integrated luminosity at 8 TeV and assuming a branching fraction BR$(\hc \rightarrow \tau \nu) = 100 \%$,  the null search results at ATLAS imply an upper bound on the production cross section $\sigma(pp \rightarrow \hc tb)$  between 0.9 pb and 0.017 pb~\cite{TheATLAScollaboration:2013wia} for charged Higgs masses between 180 GeV and 600 GeV.     When interpreting in the $m_h^{\rm max}$ scenario of the MSSM, $\tan\beta$ above 47 to 65 is excluded for $m_{\hc}$ between 230 GeV and 310 GeV. The CMS results~\cite{CMS_taunu} are very similar, which are slightly better for low $m_{H^\pm}$ and slightly worse for large $m_{H^\pm}$.
 
\begin{figure}[h!]
\centering
	\includegraphics[scale=1.0]{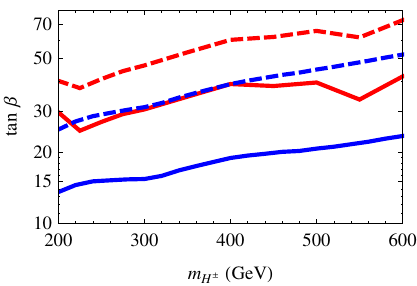}
\caption{The recast of the current ATLAS 95\% CL exclusion limits (solid red curves)~\cite{TheATLAScollaboration:2013wia} with 19.5 fb$^{-1}$ integrated luminosity and the projected $5\sigma$ reach (solid blue curves)~\cite{ATLAS-PUB} with 100 fb$^{-1}$ integrated luminosity at the 14 TeV LHC for the process $pp\to \hc tb \to(\tau\nu)(bbjj)$ in the context of the Type II 2HDM.    Also shown in dashed curves are the reduced limits with the opening of $\hc \rightarrow \A \w$, with $m_{\A}=70$ GeV, $\sba=$ 1 and ${\H}$ decoupled.     
 }
\label{fig:atlascompare}
\end{figure}

As  demonstrated in Fig.~\ref{fig:HpBR}, the conventional decay modes $\tau\nu$ and $cs$ would be highly suppressed in regions of parameter space where the exotic decay modes $\hc \rightarrow \A\w/H\w$ open. In Fig.~\ref{fig:atlascompare}, we recast the current 95\% C.L. exclusion limits (solid red curve)~\cite{CMS_taunu} and future projection of 5$\sigma$ discovery (solid blue curve)~\cite{ATLAS-PUB} with 100 fb$^{-1}$ integrated luminosity at the 14 TeV LHC for the process $pp\to \hc tb \to(\tau\nu)(bbjj)$ in the context of the Type II 2HDM.   The dashed curves show the reduced reach when $\hc \to \A \w$ opens up, shown here for the parameter choice $m_A=$  70 GeV, $\sba=$ 1, and with the $\H$ decoupled.  The  inclusion of the exotic decay modes thus substantially weakens the current and future limits.  

There have been  other theoretical studies on the charged Higgs detectability at the LHC.   The authors of \cite{THEO_SingleTop} analyzed the possibility of observing light charged Higgs decay $\hc \rightarrow \tau \nu$ via the single top production mode. The possibility of the $\hc \rightarrow \mu \nu $ decay with a light charged Higgs produced via top decay in top pair production has been investigated in~\cite{Hashemi:2011gy}. Furthermore the decay of a heavy charged Higgs into $tb$ has been studied, considering  charged Higgs production via $q q^\prime \rightarrow H^\pm$ \cite{Hashemi:2013raa},  $\hc tb$ associate production \cite{THEO_TH_TB} and $W^{\mp} \hc$ associate production \cite{THEO_WH_TB}.

Furthermore, the authors of \cite{Maitra:2014qea} studied electroweak charged Higgs boson pair production with the charged Higgses decaying into a $W$ boson and a very light [$m_{\phi} = \mathcal{O}(eV)$] neutral scalar which decays invisibly. A search strategy for $\hc \to \h \w$ for a SM-like $\h$ using the $\hc W^{\mp}$ production mode has been  suggested by the authors of \cite{Basso:2012st} and analysed in the context of CP-violating Type-II 2HDM.   This study considers both electroweak production and the production via the decay of heavy scalars, if it is kinematically allowed. Charged Higgs production via the decay of a heavy scalar $pp \rightarrow H \rightarrow  W \hc$ with $\hc \to A \w$ was investigated in \cite{Dermisek:2013cxa}.  
  
The $\hc  tb$ associated production with $\hc\to H\w \to bb\w$ has been analyzed in early studies \cite{ATLAS_EARLY_STUDY, Assamagan:2002ne}.   While Ref.~\cite{ATLAS_EARLY_STUDY} concluded that the $\hc \to \h \w / \H \w$ is not promising in MSSM searches, the authors of \cite{Assamagan:2002ne} found that this channel is indeed promising in NMSSM.   However, neither paper considers the possibility of analyzing this channel with the $\tau\tau$ mode. In particular, the $\tau\tau$ mode allows two same sign lepton signature with the accompanying leptonic decay of $W$~\cite{Khachatryan:2014qwa}, which leads to a better reach than the existing studies of the  $H/A\to bb$ channel.  Therefore, in our study, we analyze the discovery and exclusion prospects in both $\hc\to\A\w/H \w\to bb\w$ and $\hc\to\A\w/H \w\to \tau\tau\w$ channels.

 Our study also assumes the existence of a light neutral Higgs $A/H$, which has been constrained by the $A/H \rightarrow \tau\tau$ searches at the LHC~\cite{Khachatryan:2014wca,Aad:2014vgg}, in particular, for $m_{A/H}>90$ GeV and relatively large $\tan\beta$.  No limit, however, exists for $m_{A/H}<90$ GeV due to the difficulties in the identification of the relatively low $p_T$ taus and the overwhelming SM backgrounds for low $p_T$ leptons and $\tau$-jets.   Furthermore, LEP limits~\cite{LEP_Higgs} based on $VH$ associated production do not apply for the CP-odd $A$ or the non-SM like CP-even Higgs.  LEP limits based on $AH$ pair production also do not apply as long as $m_A+m_H>208$ GeV.  Therefore, in our analyses below, we choose the daughter neutral Higgs mass to be 70\footnote{The mass of 70 GeV was choose to be above the $h_{\rm SM} \to AA$ threshold to avoid significant deviations of the 126 GeV SM-like Higgs branching fractions from current measurements.}, 126, and 200 GeV  to represent the cases with a light, SM-like, and a heavy neutral Higgs respectively.

\section{Collider analysis}
 \label{sec:analysis}
 
 \subsection{Signal process}
 \label{sec:signal}

In our analysis we  study  the associated production $pp \rightarrow  \hc tb$ in which the charged Higgs boson decays into a neutral Higgs ($A$ or $H$) and a $W$.   The  dominant leading order Feynman  diagrams contributing to this production are shown  in Fig. \ref{fig:signal_diag}~\cite{Dittmaier:2009np}.   For large charged Higgs masses,  diagrams (a) and (b)  dominate while for smaller charged Higgs masses, top pair production in panel (c)  with the decay of one (possibly offshell) top into a charged Higgs dominates\footnote{All possible production diagrams are taken into account for event generation.}.  The exclusion and discovery reach in $\sigma\times\rm{BR}$  obtained in this section will cover the entire kinematically possible mass range.  When interpreting the results in the Type II 2HDM in Sec.~\ref{sec:implication}, we  focus on the high mass region: $m_{\hc} > m_t$.  For  the low mass range where  the $t\bar{t}$ production dominates, the bounds are usually translated into limits on the branching fraction BR$(t \to \hc b)$ \cite{Kling_future}.

\begin{figure}[h!]
 \centering
 	\minigraph{3cm}{-0.in}{(a)}{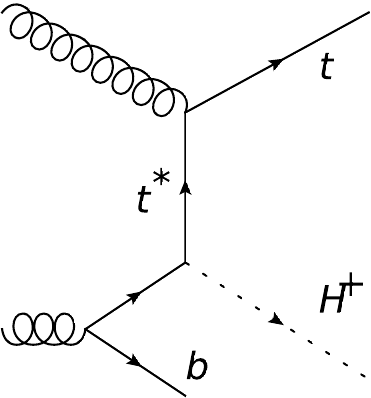} \hspace{0.5 cm}
  	\minigraph{4.5cm}{-0.in}{(b)}{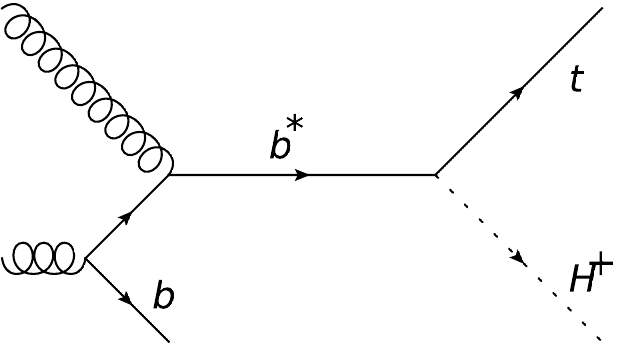} \hspace{0.5 cm}
 	\minigraph{4.5cm}{-0.in}{(c)}{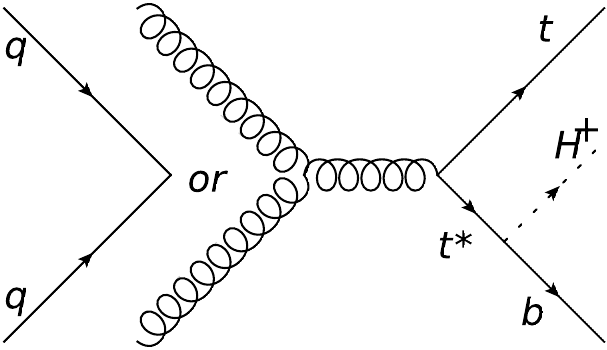}
\caption{Dominant $t$-channel (a), $s$-channel (b) and $t\bar{t}$-like (c) diagrams contributing to heavy quark associated charged Higgs production \cite{Dittmaier:2009np}. }
\label{fig:signal_diag}
\end{figure}

In principle the neutral Higgs boson can either be CP-even (denoted by $H$) or CP-odd (denoted by $A$). In the analysis that follows, we use the decay $\hc \rightarrow A \w$ as an illustration. Since we do not make use of angular correlations, the bounds obtained for $\hc \rightarrow A \w$ apply to  $\hc \rightarrow H \w$ as well.

The neutral Higgs boson itself will further decay. We only look at the fermionic decays $A \rightarrow bb, \tau\tau$.  While the $bb$ case has the advantage of a large branching fraction $\text{BR}(A \rightarrow bb)$, the $\tau\tau$ case has  less SM backgrounds   and therefore leads to a cleaner signal. We study both leptonic and hadronic $\tau$ decays and consider the three cases: $\tau_{had}\tau_{had}$,  $\tau_{lep}\tau_{had}$ and  $\tau_{lep}\tau_{lep}$.  The $\tau_{lep}\tau_{had}$ case is particularly promising since we can utilize the same sign dilepton signal with the leptons from $W$ decay and from $\tau$ decay.  Exotic decays of $A/H$  into pairs of vector bosons or other Higgs bosons will most likely be suppressed or have a very complex final state.  Since the top quark decays to $bW$, the final state contains two $W$ bosons. To reduce the backgrounds, in our analysis we assume one of these two $W$ bosons decays leptonically, with the other $W$ decaying hadronically. 

We use Madgraph 5/MadEvent v1.5.11 \cite{Alwall:2011uj} to generate our signal and background events.  These events are passed to Pythia v2.1.21 \cite{PYTHIA}  to simulate initial and final state radiation,  showering and hadronization. The events are further passed through Delphes 3.07 \cite{Ovyn:2009tx} with the Snowmass combined LHC detector card \cite{snowmassdetector} to simulate detector effects. The discovery reach and exclusion bounds have been determined using the program  RooStats~\cite{Moneta:2010pm}  and theta-auto \cite{thetaauto}.

In this section, we will present model \emph{independent} limits on the $\sigma\times\rm{BR}$ for both 95\% C.L.  exclusion and 5$\sigma$ discovery for both possible final states $\tau\tau bbWW$ and $bbbbWW$. For the signal process, we generated event samples at 14 TeV LHC for $pp \rightarrow  \hc tb \rightarrow  A \w tb$ with the daughter particle mass fixed at $m_A= 70, 126, 200$ GeV to represent the cases with a light, SM-like, and a heavy Higgs respectively. For each case, we vary the parent particle mass $m_{\hc}$ in the range 150 $-$ 600 GeV.


\subsection{$A \rightarrow \tau\tau$ mode}
\label{sec:tautau}
 
 We start our analysis by looking at the channel $pp \rightarrow  \hc tb \rightarrow \A\w tb\rightarrow \tau\tau bbWW $.    We only require to identify one $b$ jet  from top decay. We do not require to find the $b$ jet  produced in association with the charged Higgs since it  is likely to be soft.   As mentioned above,  we will distinguish three cases depending on how the taus decay:  
 \begin{itemize}
 \item \textbf{Case A:} Both taus decay hadronically.
  \item \textbf{Case B:} One tau decays hadronically, and the other tau decays leptonically.
   \item \textbf{Case C:} Both taus decay leptonically.
\end{itemize}
For the two $W$ bosons, we require one decay leptonically and the other decay hadronically.  
The dominant SM background for this final state is semi- and fully leptonic (where leptonic includes decaying into $\tau$) $t\bar{t}$ pair production, which we generate with up to one additional jet. We also take into account $tt\tau\tau$ production,  where the taus come from the decay of a boson $Z/H/\gamma^*$. Furthermore we include $W\tau\tau$ production with up to two additional jets (including $b$ jets)  and $WW\tau\tau$ production with up to one additional jet  (including $b$ jet), where the taus are produced in the decay of a boson $Z/H/\gamma^*$. We   ignored the subdominant backgrounds from single vector boson production, $WW$, $ZZ$, single top production, as well as multijet QCD Background.   Those backgrounds are either small or can be sufficiently suppressed by the cuts imposed.

We apply the following cuts to extract the signal from the backgrounds:
\begin{enumerate}
	\item \textbf{Identification cuts:} \\
\textbf{Case A:} One lepton $\ell=e$ or $\mu$ , one or two $b$ jets, two $\tau$ tagged jets and at least two untagged jets:
\begin{equation}
\begin{aligned}
n_{\ell}=1,  \ n_b=1,2, \ n_{\tau}& =2, \ n_{j}\ge 2.
\end{aligned}
\label{cut_t1A}
\end{equation}
We require that the $\tau$ tagged jets have opposite charge.

\textbf{Case B:} Two leptons, one or two $b$ jets, one $\tau$ tagged jet and at least  two untagged jets:
\begin{equation}
\begin{aligned}
n_{\ell}=2,  \ n_b=1,2,  \ n_{\tau}& =1, \ n_{j}\ge 2. 
\end{aligned}
\label{cut_t1B}
\end{equation}
We require that both leptons have the same sign, which is opposite to the sign of the $\tau$ tagged jet.

\textbf{Case C:} Three leptons, one or two $b$ jets, no $\tau$ tagged jet and at least  two untagged jets: 
\begin{equation}
\begin{aligned}
n_{\ell}=3,  \ n_b=1,2,  \ n_{\tau}& =0, \ n_{j}\ge 2. 
\end{aligned}
\label{cut_t1C}
\end{equation}

 We adopt the following selection cuts for the identification of leptons, $b$ jets and jets. 
\begin{equation}
\begin{aligned}
 |\eta_{\ell,b,\tau} |<2.5,\ |\eta_{j} | &<5, \ p_{T;\ell_1,j,b} > 20\ {\rm GeV} \ \text{and} \  p_{T;\ell_2} > 10\ {\rm GeV}, 
\end{aligned}
\label{cut_t1}
\end{equation}
where  $\ell_{1,2}$ refer to the hardest and  the sub-leading lepton.   For jet reconstruction, the anti-$k_T$ jet algorithm with $R=$ 0.5 is used.    

	\item \textbf{Two $W$ candidates:} Our analysis assumes that one $W$ decays leptonically and the other decays hadronically. We look for the combination of two untagged jets that gives an invariant mass closest to the $W$ mass and reconstruct the jets to form the hadronic $W_{had}$. The momentum of the neutrino coming from the leptonic $W$ decay is determined using the missing transverse momentum and imposing the mass conditions  \cite{Aad:2012ux}. Using the momenta of the reconstructed neutrino and the lepton, the momentum of the leptonic $W_{lep}$ can be deduced. In cases B and C which contain more than one lepton, the hardest lepton is used for $W$ reconstruction. In these cases the neutrino reconstruction will be relatively poor since there is additional missing energy from the $\tau$ decay.
	
	\item \textbf{Top candidate:} We look for the combination of the $b$ tagged jet and a reconstructed (either leptonic or hadronic) $W$ that gives an invariant mass closest to the top mass and combine them to form the top candidate $t$. 
	
	\item \textbf{Neutral Higgs candidate ($H$):} The $\tau$ jets (case A), the $\tau$ jet and the softer lepton (case B) or the two softer leptons (case C) are combined to form the neutral Higgs candidate. In cases B and C the Higgs reconstruction will be relatively poor for reasons mentioned above which in turn forces us to employ more relaxed mass cuts (see below).
	
	\item \textbf{Charged Higgs candidate ($H^{\pm}$):} The Higgs candidate  and the $W$ candidate not used for the top reconstruction are combined to form the charged Higgs candidate $H^{\pm}$. 	
	
	\item \textbf{$m_{\tau\tau}$ versus $m_{\tau\tau W}$:} We require the ditau mass $m_{\tau\tau}$ to be close to the daughter Higgs mass $m_A$ and the mass of the two taus and the $W$ ($m_{\tau\tau W}$) to be close to the parent Higgs mass $m_{\hc}$. The two masses are correlated, i.e., if we underestimate $m_{\tau\tau}$ we also underestimate $m_{\tau\tau W}$. To take this into account we apply a two-dimensional cut: 
\begin{equation}
\begin{aligned}
(1-\Delta  - w_{\tau\tau})\cdot m_{A} &< \;  m_{\tau\tau} < (1-\Delta  + w_{\tau\tau})\cdot m_{A},   \\
\frac{m_{A}}{E_A}(m_{\tau\tau W}-  m_{\hc} - w_{\tau\tau W} )< \ & m_{\tau\tau}  -m_A < \frac{m_{A}}{E_A}(m_{\tau\tau W } -m_{\hc} + w_{\tau\tau W} ). 
\end{aligned}
\label{cut_tau3}
\end{equation}	
 Here $w_{\tau\tau}$ = 0.225 (case A) or 0.25 (cases B and C) is the width of the ditau mass window. 
Note that the slightly shifted reconstructed Higgs mass $m_{\tau\tau}$ around $(1-\Delta) m_A$ instead of $m_A$  is due to the reconstruction of the $\tau$ using a jet with a small size of $R=$ 0.5 or a lepton. We use $\Delta$ = 0.3 (case A), 0.4 (case B) and 0.66 (case C).  The second condition describes two lines going through the points $(m_{\hc} \pm w_{\tau\tau W }, m_A)$ with slope $\frac{m_{A}}{E_A}$ where $E_A$ is the energy of the neutral Higgs in the rest frame of the charged Higgs\footnote{We choose $E_A = \frac{m_{\hc}^2 + m_A^2 - m_W^2}{2 m_A}$, which is the energy of $\A$ in the rest frame of the charged Higgs.  The slope in Eq.~(\ref{cut_tau3}) can be motivated by relativistic kinematics and works well even when the charged Higgs is not produced at rest. }. We choose a width for the $m_{\tau\tau W}$ peak of $w_{\tau\tau W} = 0.2  m_{\hc}$, based on the  theoretical decay width estimation of $w_{\hc} \sim 0.1 m_{\hc}$   as well as detector resolutions. The effectiveness of this cut is shown in Fig.~\ref{fig:2D_mhvsmc} for $m_{\hc}=240$ GeV and $m_{A}= 70$ GeV in case A, with two horizontal lines indicating the $m_{\tau\tau}$ range and two slanted lines indicating the $m_{\tau\tau W}$ range as given in Eq.~(\ref{cut_tau3}).
\end{enumerate}

 \begin{figure}[h!]
 \centering
	\includegraphics[scale=0.35]{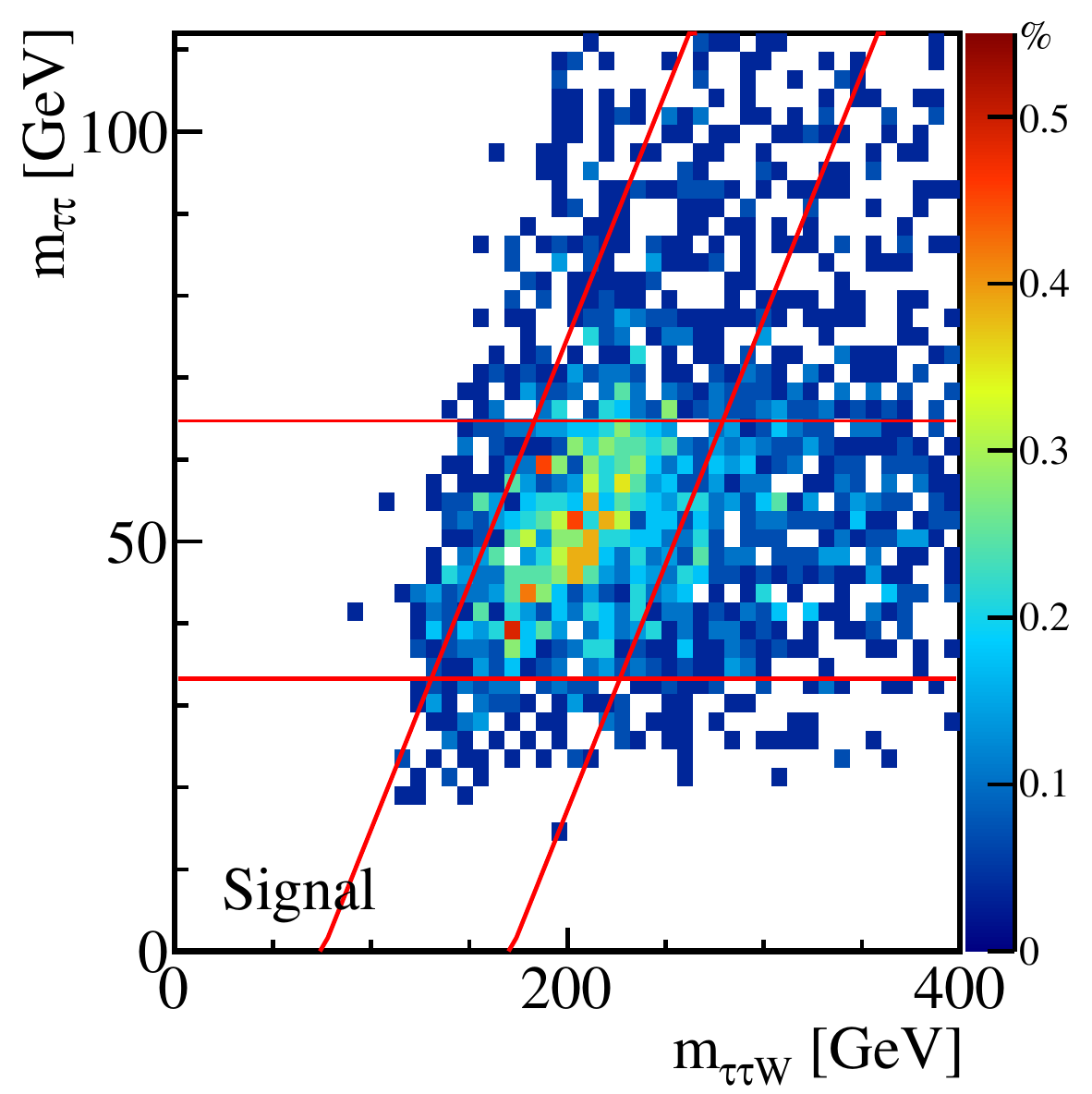}
	\includegraphics[scale=0.35]{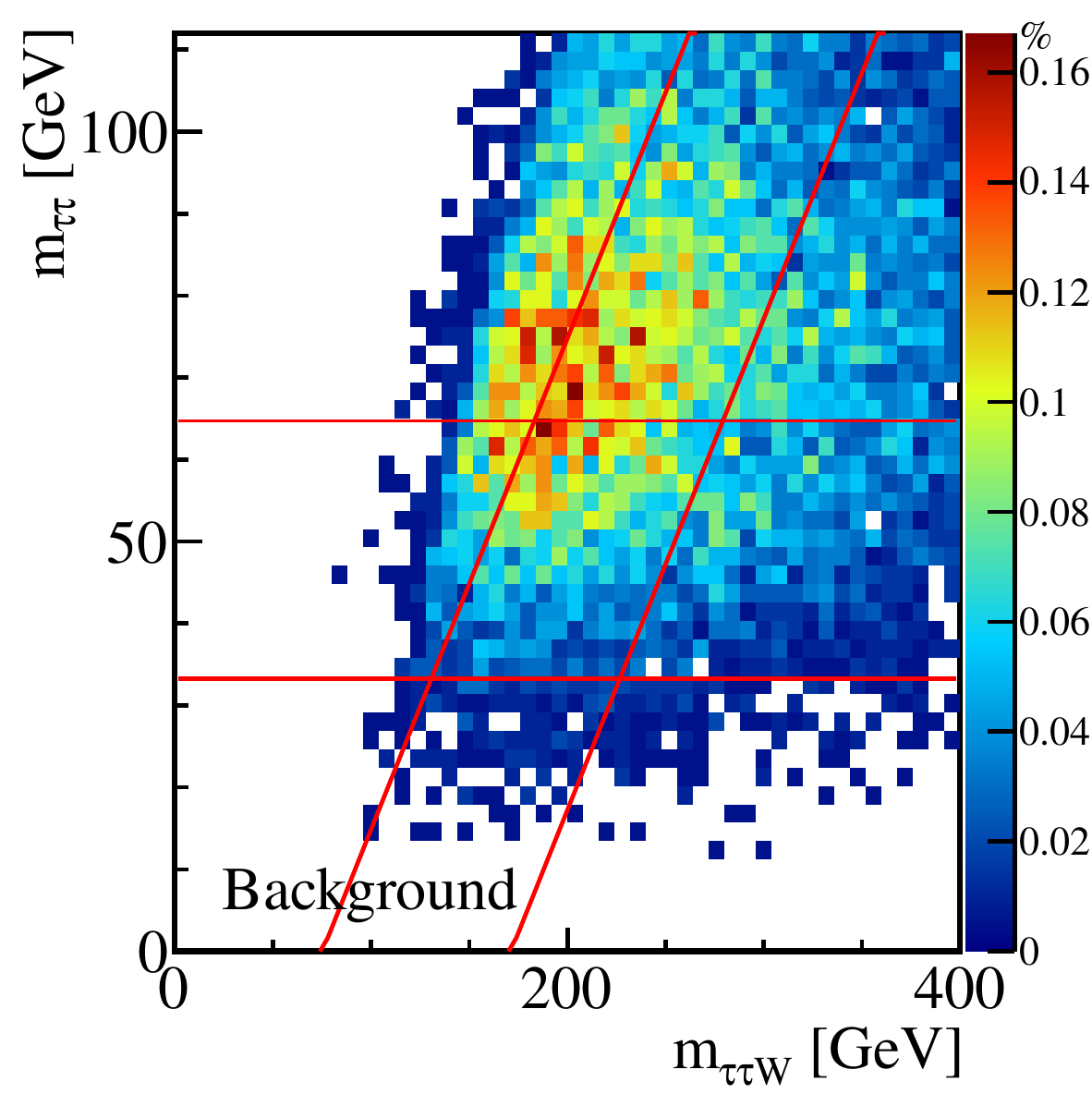}
\caption{Normalized distribution (in percent as given by the color code in the panel along the $y$-axis) of $m_{\tau\tau}$ versus  $m_{\tau\tau W}$ for the signal (left) and the backgrounds (right) assuming $m_{\hc}=240$ GeV and $m_{A}= 70$ GeV for case A. Two horizontal lines indicate the $m_{\tau\tau}$ range and two slanted lines indicate the $m_{\tau\tau W}$ range, as given in Eq.~(\ref{cut_tau3}).   }
\label{fig:2D_mhvsmc}
\end{figure}

No mass cuts are applied for the reconstructed $W$ and $t$ candidates since both signal and the dominant backgrounds contain a top quark and an additional $W$ boson. In Table \ref{tab:semi_tata}, we show the signal and background cross sections with cuts for a signal benchmark point of $M_{\hc} = 240$ GeV and $m_A =  70$ GeV at the 14 TeV LHC. The first row shows the total cross section before cuts calculated using MadGraph.    The following rows show the cross sections after applying the identification cuts and mass cuts for all three cases as discussed above. We have chosen a nominal value for $\sigma \times BR( p p  \rightarrow \hc t b \rightarrow \tau \tau b bWW )$ of 100 fb\footnote{For the  Type II 2HDM the cross section for $m_{\hc}=240$ GeV is typically in the range of $\sigma(pp\to \hc tb)=$ 0.1$-$1.5 pb (see Fig. \ref{fig:SigmaHC}.). Assuming a branching fraction BR$(\hc \to A\w)  = 100 \%$ and BR$(A \to \tau\tau)=10 \%$ leads to the stated $\sigma\times$ BR of around 100 fb.   } to illustrate the cut efficiencies for the signal process. The last column shows the $S/\sqrt{B}$ value  for an integrated luminosity of 300 \ifb.

\begin{table}[h]
\centering
\resizebox{14cm}{!} {
\begin{tabular}{ |c | r |r r r |r r|}  \hline
Cut 		 						&Signal [fb] 	&$t\bar{t}$ [fb] 		&$t\bar{t}\tau\tau$ [fb]	&$W(W)\tau\tau$ [fb]	&$S/B$	&$S/\sqrt{B}$		\\
\hline
$\sigma$    							&100 		&$6.3 \cdot 10^5$		&  247   		 	& 2000    		&-	&		-	\\
\hline
 A: Identification [Eq.(\ref{cut_t1A})] 		 	&0.57 		&           22.9 		&    0.58			& 0.078     		& 0.02	&	2.04		\\
\ \  \ $m_{\tau\tau}$ vs $m_{\tau\tau W}$ [Eq.(\ref{cut_tau3})]	&0.16 		&           1.67 		&    0.054 			& 0.010     		& 0.10	&	2.20		\\
\hline
 B: Identification [Eq.(\ref{cut_t1B})] 			&0.47 		&           0.35 		&    0.697 			& 0.073     		& 0.42	&	7.81		\\
 \ \ \  $m_{\tau\tau}$ vs $m_{\tau\tau W}$ [Eq.(\ref{cut_tau3})]&0.15 		&           0.043 		&    0.104 			& 0.018   		& 0.94	&	6.67		\\
\hline
 C: Identification [Eq.(\ref{cut_t1C})] 	 		&0.48 		&           2.35 		&    5.11 			& 0.059     		& 0.06	&	3.05		\\
 \ \  \ $m_{\tau\tau}$ vs $m_{\tau\tau W}$ [Eq.(\ref{cut_tau3})]&0.15 		&           0.56 		&    0.56 			& 0.010     		& 0.13	&	2.54		\\
\hline
\end{tabular}
}
\caption{Signal and background cross sections with cuts for the signal benchmark point $m_{\hc}$ = 240 GeV   and $m_A$ =  70 GeV at the 14 TeV LHC.  We have chosen a nominal value for $\sigma \times {\rm BR}(pp \rightarrow \hc tb \rightarrow \tau\tau bb WW)$ of 100 fb to illustrate the cut efficiencies for the signal process.  The last column of $S/\sqrt{B}$ is shown for an integrated luminosity of ${\cal L}=300\  {\rm fb}^{-1}$.  }
\label{tab:semi_tata}
\end{table}

We can see that the dominant background contributions are $t\bar{t}$ (case A) and $t\bar{t}\tau\tau$ (cases B and C) while the vector boson backgrounds do not contribute much. It turns out that case B, in which one $\tau$ decays leptonically and the other $\tau$ decays hadronically, gives the best reach. This is because the same sign lepton signature can reduce the $t\bar{t}$ background sufficiently. This analysis is sensitive to the tagging and misidentification rate of the $\tau$ tagger. Most of the top pair background, especially in case A, includes mistagged $\tau$ jets. We assume a tagging rate of $\epsilon_{tag}=60\%$ and a mistagging rate of $\epsilon_{miss}=0.4\%$ as suggested in \cite{snowmassdetector}.  A better rejection of non-$\tau$ initiated jets would increase the significance of this channel.

 \begin{figure}[h!]
 \centering
 	\includegraphics[scale=1]{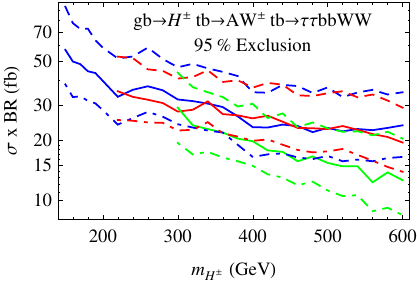}
 	\includegraphics[scale=1]{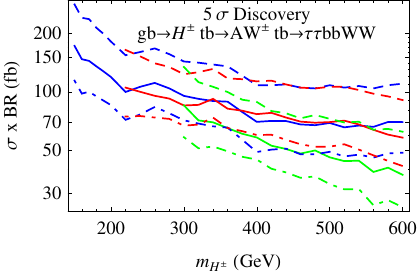}
\caption{The 95\% C.L. exclusion (left) and 5$\sigma$ discovery (right) limits for $\sigma \times \text{BR}(pp \to \hc t b \to \tau\tau bb WW )$ for $m_A$ = 70 GeV (blue), 126 GeV (red), and 200 GeV (green) at the 14 TeV LHC. We have combined all three cases of tau decays. The dashed, solid and dot-dashed lines correspond to an integrated luminosity of 100, 300 and 1000 fb$^{-1}$, respectively. Here, we have assumed a 10\% systematic error on the backgrounds. These results are equally applicable to the $\hc \to H \w$ process for the same parent and daughter Higgs masses.    
}
\label{fig:limits_tautau}
\end{figure}

In Fig.~\ref{fig:limits_tautau}, we display the results at the 14 TeV LHC for 95\% C.L. exclusion (left panel) and 5$\sigma$ discovery (right panel)  limits for $\sigma \times {\rm BR} (pp\to\hc t b \to \tau\tau bb WW)$, which applies for $\hc \to H \w$ as well with $m_A$ replaced by $m_H$.  We have combined all three cases of tau decays.   The blue, red, and green curves   correspond to the daughter particle being  70 GeV, 126 GeV, and 200 GeV,  respectively. For each mass, we have displayed the results for three luminosities: 100 fb$^{-1}$ (dashed),  300 fb$^{-1}$ (solid),  and 1000 fb$^{-1}$ (dot-dashed), with 10\% systematic error included \cite{thetaauto}. Due to the small number of events, the statistical error dominates in this channel and therefore higher luminosities lead to a better reach. Better sensitivity is achieved for larger $m_{\hc}$ since the mass cuts on $m_{\tau\tau}$ and $m_{\tau\tau W}$ have  a more pronounced effect on the SM backgrounds for larger masses.    

The $m_{\tau\tau}$ distribution for the dominating $tt$ backgrounds peaks around higher masses $m_{\tau\tau} \approx $ 70 - 200 GeV and therefore the background rejection efficiency for $m_{\tau\tau} \approx$  70 GeV is high compared to the cases with larger daughter particle masses. On the other hand a small daughter Higgs mass causes the taus to be either soft (low $m_{\hc}$) or collimated (high $m_{\hc}$) and decreases the identification efficiency compared to higher daughter particles masses.  Taking into account these two effects, the limits do not change significantly for $m_A$ being 70 GeV or 125 GeV.  The limit for $m_A=200$ GeV is  better by about a factor of 1.5.

The limit, however, gets  slightly worse for  the $m_A=  70$ GeV case when  $m_{\hc} \gtrsim $  500 GeV (blue curves).  This is due to the decrease of the signal cut efficiency for a highly boosted  daughter particle with two collimated  $\tau$ jets.  For the interesting case where the daughter particle is  70 GeV, it is seen that the exclusion limits for a 300 fb$^{-1}$ collider fall from about  60 fb  for $m_{\hc}$ of $150$ GeV, to less than 25 fb for a 500 GeV charged Higgs.    The 5$\sigma$ discovery limits are about a factor of 3$-$4 higher.  

We reiterate here that these exclusion and discovery limits are completely model independent. Whether or not discovery/exclusion is actually feasible in this channel should be answered within the context of a particular model, in which the theoretically predicted cross sections and branching fractions can be compared with the exclusion or discovery limits.   We will do this in Sec.~\ref{sec:implication} using the Type II 2HDM as a specific  example.


 \subsection{$A \rightarrow bb$ mode}
\label{sec:ana_bb}

We now turn to the channel $pp \rightarrow  \hc tb \rightarrow bbbbWW$, with one $W$ decaying leptonically and the other decaying hadronically.  The dominant SM backgrounds for this final state are semi- and fully leptonic top pair production, which we generate with up to one additional jet. We also take into account $ttbb$ production where the two bottom jets  either come from the decay of a boson $Z/H/\gamma^*$ or are produced through gluon splitting. We have ignored the subdominant backgrounds including  $V+$jets, $VV+$jets or $VVV$+jets, single top production, as well as multijet QCD Background. These backgrounds either have small production cross sections, or can be sufficiently suppressed by the cuts imposed.

Much of the analysis for this case is similar to the $\tau\tau$ case described above. We apply the following cuts to identify the signal from the backgrounds:
\begin{enumerate}
	\item \textbf{One lepton, three or 4 $b$ jets, at least  two untagged jets:}  
\begin{equation}
\begin{aligned}
n_{\ell}=1,  \ n_{b}& =3,4,  \ n_{j}\ge 2 \ {\rm with}\ \\
 |\eta_{\ell,b} |<2.5,\ |\eta_{j} | &<5, \ p_{T,\ell,j,b} > 20\ {\rm GeV.}
\end{aligned}
\label{cut_b1}
\end{equation}

	\item \textbf{Two $W$ candidates and one top candidate:} Similar to that in Sec.~\ref{sec:tautau}. For top reconstruction,  we look for the combination of a $b$ tagged jet and a reconstructed $W$ that gives an invariant mass closest to the top mass.

	\item \textbf{Neutral Higgs candidate ($A$):} The remaining $b$ jets are combined to form the Higgs candidate $A$ with mass $m_{bb}$. 
	
	\item \textbf{Charged Higgs candidate ($H^{\pm}$):} The  Higgs candidate and the $W$ candidate not used for the top reconstruction  are combined to form the charged Higgs candidate $H^{\pm}$ with mass $m_{bbW}$. 	
	
	\item \textbf{$m_{bb}$ versus $m_{bb W}$:} There is no   Higgs mass shift $\Delta$ as in the $\tau\tau$ case since there is no missing energy carried away by neutrinos from tau decay anymore.   Our 2-D cuts are thus modified as follows: \begin{equation}
\begin{aligned}
(1 -  w_{bb})\cdot m_{A} &< \;  m_{bb} < (1  + w_{bb})\cdot m_{A},   \\
\frac{ m_A}{E_A}(m_{bb W}-  m_{\hc} - w_{bb W} )< \ & m_{bb}  -m_A < \frac{ m_A}{E_A}(m_{bb W } -m_{\hc} + w_{bb W} ).
\end{aligned}
\label{cut_b3}
\end{equation}	
The mass window chosen is slightly   tighter due to a better mass reconstruction in the $bb$ case: $w_{bb} = 0.2$ and $w_{bb W} = 0.175 m_{\hc}$.

\end{enumerate}
 In Table \ref{tab:semi_bb1}, we present the cross sections after the individual cuts are imposed sequentially. We take a nominal signal cross section of 1000 fb to illustrate the efficiency of the chosen cuts. Since the expected number of events is large, the systematic uncertainty will dominate and a large ratio $S/B$ is desired. Although $S/\sqrt{B}$ does not improve using the mass cut , $S/B$ improves and therefore the systematic uncertainty, which dominates the overall uncertainty, decreases. The dominant background comes from top pair production. 

\begin{table}[h]
\centering
\resizebox{14cm}{!} {
\begin{tabular}{| c | r |r r |r r|}
\hline
Cut 								&Signal [fb] 	&$t\bar{t}$ [fb] 		&$t\bar{t}bb$ [fb]	&$S/B$	&$S/\sqrt{B}$	\\
\hline
$\sigma$   							&1000		&$6.5 \cdot 10^5$		&11310		  	&-	&		-	\\
\hline
  Identification [Eq.(\ref{cut_b1})] 		 		&13.3 		&         903 			&   143			& 0.012	&	7.1		\\
  $m_{bb}$ vs $m_{bb W}$ [Eq.(\ref{cut_b3})]			&0.83 		&          28 			&   3.8 		& 0.026	&	2.5		\\
\hline
\end{tabular}
}
\caption{Signal and background cross sections with cuts for the signal benchmark point $m_{\hc}$ = 240 GeV and $m_A$ =  70 GeV at the 14 TeV LHC. We have chosen a nominal value for $\sigma \times BR(pp \rightarrow \hc tb \rightarrow b bbb WW)$  of 1000 fb to illustrate the cut efficiencies for the signal process.  The last column of $S/\sqrt{B}$ is shown for an integrated luminosity of ${\cal L}=300\  {\rm fb}^{-1}$. }
\label{tab:semi_bb1}
\end{table}

\begin{figure}[h!]
 \centering
 	\includegraphics[scale=1]{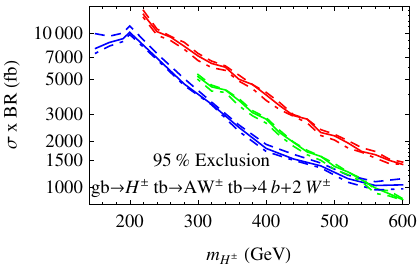}
 	\includegraphics[scale=1]{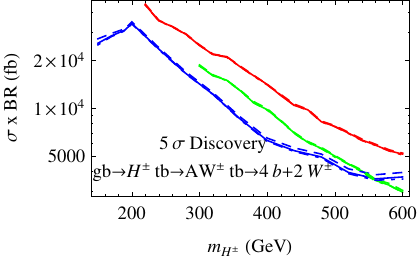}
\caption{The 95\% C.L. exclusion (left) and 5$\sigma$ discovery (right) limits for $\sigma \times \text{BR}(pp \to \hc t b \to bb bb WW )$ for $m_A$ =  70 GeV (blue), 126 GeV (red), and 200 GeV (green) at the 14 TeV LHC. The dashed, solid and dot-dashed lines correspond to an integrated luminosity of 100, 300 and 1000 fb$^{-1}$ respectively. Here, we have assumed a 10\% systematic error on the backgrounds.   }
\label{fig:limits_bb}
\end{figure}

In Fig. \ref{fig:limits_bb}, we show the 95\% C.L. exclusion and 5$\sigma$ discovery reach in $\sigma \times \text{BR}(pp \to \hc t b \to bbbbWW)$ for the 14 TeV LHC. The general feature of these plots follows that of Fig. \ref{fig:limits_tautau}, particularly with highly boosted daughter particles making $b$ identification more  challenging, as shown by the flattening  of the blue curves for  70 GeV daughter particle mass when  $m_{H^\pm} \gtrsim$  550 GeV.   Unlike  the $\tau\tau$ case, different luminosities do not change the   limits significantly as the errors on the backgrounds are dominated by systematic uncertainties. Thus, in our analysis, we have chosen a uniform 10\% systematic error on the backgrounds.  With the possible reduction of systematic errors in the future, the cross section  limits can be improved. For example, a 5\% systematic error would lead to the cross section limits   improved by about a factor of 2.    The exclusion limits are lowest for small $m_A$ =  70 GeV since the dominating $t\bar{t}$   background peaks around $m_{bb} \approx$  70$-$200 GeV and therefore the background rejection efficiency for $m_{bb} \approx$  70 GeV is high.   The improvement of the sensitivity for the $m_A= 70$ GeV case when $m_{\hc}<200$ GeV is due to the suppression of the $t\bar{t}$ background with the $m_{bbW}$ cut.

Compared to the $\tau\tau$ case, the $\sigma \times \text{BR}$ reach in the $bb$ case is worse due to significantly higher SM backgrounds. For the  70 GeV daughter particle case with 300 fb$^{-1}$, the exclusion limit varies from about 10 pb for a parent mass of 200 GeV to about 1.5 pb for 500 GeV. Thus, given the typical ratio of BR $(A/H \to bb) : {\rm Br}(A/H\to \tau\tau) \sim 3m^2_b/m^2_{\tau}$, we conclude that the reach in the $bb$ case is much worse than that in  the $\tau\tau$ case for all masses.

\section{Implication for the Type II 2HDM}
 \label{sec:implication}

The discussion thus far has been completely model independent, and the discovery and exclusion limits displayed in Figs.~\ref{fig:limits_tautau} and \ref{fig:limits_bb} apply to any model in which    $\hc \rightarrow \A\w/H\w$  occurs. In this section, we will analyze the feasibility of this channel at the 14 TeV LHC in the context of the Type II 2HDM. 

\subsection{Cross section and branching fractions}
In the Type II 2HDM,  one Higgs doublet $\Phi_1$ provides masses for  the down-type quarks and charged leptons, while the other Higgs doublet $\Phi_2$ provides masses for the up-type quarks.  The couplings of  the  CP-even Higgses $\h$, $\H$ and the CP-odd Higgs $\A$ to the SM particles  can be found in Ref.~\cite{Branco:2011iw}.

The discovery of the 126 GeV SM-like Higgs imposes restrictions on the couplings and masses of the various Higgses in the 2HDM, and several studies in the literature mapped  out the available parameter space after all the theoretical and experimental constraints are imposed \cite{Craig:2012vn, Chiang:2013ixa, Grinstein:2013npa, Coleppa:2013dya,  2HDM_other}. Note that the 2HDM offers two possibilities: either the $\h$ or the $\H$ could be interpreted as the observed 126 GeV resonance, and accordingly, the available parameter spaces differ. In the $\h$-126 case with $m_{12}^2=0$, we are restricted   to narrow regions with $\sba\sim\pm$ 1 with $\tan\beta$ up to 4 or an extended region in $0.55 < \sba < 0.9$ with $1.5 < \tan\beta<4$. The masses $m_{\H}, m_{H^\pm}$, and $m_{\A}$ are, however, relatively unconstrained. In the $\H$-126 case with $m_{12}^2=0$, we are restricted to a narrow region of $\sba\sim$ 0 with $\tan\beta$ up to about 8, or an extended region of $\sba$  between $-0.8$ to $-0.05$,  with $\tan\beta$ extending to 30 or higher \cite{ Coleppa:2013dya}. $m_{\A}$  and $m_{H^\pm}$ are nearly degenerate due to $\Delta\rho$ constraints.  Imposing the flavor constraints further narrows down the preferred parameter space. In what follows, we will specify the Higgs masses  for each benchmark point considered, but will display our results for all values of $\sba$ and $\tan\beta$.

Fig.~\ref{fig:SigmaHC} shows contours of  NLO $\sigma(gg\to \hc tb )$ in the $m_{\hc}-\tan\beta$ plane at the 14 TeV LHC, with  values  taken from the LHC Higgs Working Group \cite{LHC-WorkingGroup}\footnote{The NLO  cross sections are available only for $m_{\hc}\geq$200 GeV. Thus, for $m_{\hc}$ less than this value, we simply using the leading order numbers calculated using FeynHiggs~\cite{Feynhiggs}.}. The production is controlled by the $\hc tb$ vertex, which is given in Eq.~(\ref{eqn:htbvertex}).    This coupling is enhanced for both small and large $\tan\beta$, due to the enhancement of the top and bottom Yukawa coupling, respectively.   Correspondingly,  the cross section can reach up to 1.5 pb for $m_{\hc}\leq$ 300 GeV for either $\tan\beta>$ 40, or $\tan\beta<$ 2. However, we note that the cross section decreases rapidly with increasing mass,  falling below 50 fb   in most regions of  $m_{\hc}>$ 400 GeV. This makes the charged Higgs search challenging in the high mass regions unless we get a  particularly clean signal with minimal backgrounds.       

\begin{figure}[h!]
\centering
	\includegraphics[scale=1]{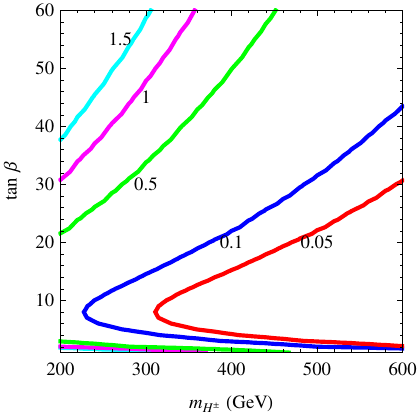}
\caption{ Contours of   NLO $\sigma(pp\to  \hc tb)$ (in pb)  in the $m_{\hc}-\tan\beta$ plane at the 14 TeV LHC for the Type II 2HDM.   }
\label{fig:SigmaHC}
\end{figure}


The results of Sec.~\ref{sec:analysis}, in principle, could be interpreted within the context of three processes: $\hc\rightarrow\A\w$, $\hc\to\h\w$, and $\hc\to\H\w$. The decay width of the first of these is independent of $\sba$,   while decay to $\h\w$ or $\H\w$ is proportional to $\cos(\beta-\alpha)$  or $\sin(\beta-\alpha)$.   Therefore, the decay to non-SM-like  Higgs is preferable.   In  this section, we will consider two cases for illustration: i) $\hc\to A\w$ for the $\h$-126 case with   $\H$ decoupled and ii) $\hc\to \h\w$ for the $\h$-126 and $\H$-126 cases with $\A$ decoupled.  We do not consider the decay $\hc\to \H\w$ as its reach is similar to the $\hc\to \A\w$ channel  in the $\h$-126 case while being suppressed in the $\H$-126 case.  We do not consider the decay $\hc\to \A\w$  in the $\H$-126 case since the reach is always worse that that in the $\h$-126 case due to competition from the $\hc \rightarrow \h\w$ mode.

We list the specific benchmark points considered in Table~\ref{tab:classification}.     BP1 and BP2 are chosen to illustrate the reach for the $\hc\rightarrow A\w$ decay.  A smaller $m_{\hc}$ is chosen for BP1 to illustrate the effect of a larger production cross section.    BP3 and BP4 are chosen to illustrate the reach for the $\hc\rightarrow \h\w$ decay, with unsuppressed decay in BP3 ($\H$-126 case)  and suppressed decay in BP4 ($\h$-126 case) when preferred value of $\sin(\beta-\alpha)$ is considered.    Note that BP1 and BP4 admit only one exotic decay ($A\w$ for the former and $\h\w$ for the latter),  thus  representing the simplest scenario where the  reach is maximized in these two modes for the chosen $m_{\hc}$ value.  

\begin{table}[h]
\begin{center}
  \begin{tabular}{|l|c|c|c|c| }
    \hline
    $\left\{ {m_{\hc},m_{\A},m_{\h},m_{\H}}\right\}$ GeV & $\hc\to\A \w$ & $\hc\to\h \w$ & Favored Region  \\ \hline
   BP1: $\left\{ {200, 70, 126, 700}\right\}$ & \cmark & \xmark &  $\sba\approx \pm$ 1 \\ \hline
    BP2: $\left\{ {300, 126, 126, 700}\right\}$ & \cmark & \cmark & $\sba\approx\pm$ 1 \\ \hline
     BP3: $\left\{ {300, 700, 70, 126}\right\}$ & \xmark & \cmark & $\sba\approx$ 0 \\ \hline
    BP4: $\left\{ {300, 700, 126,  700}\right\}$ & \xmark & \cmark & $\sba\approx\pm$ 1 \\ \hline
   \end{tabular}
\end{center}
\caption{Benchmark points shown for illustrating the discovery and exclusion limits for  the processes $pp \rightarrow \hc tb \rightarrow \A\w/H\w tb \rightarrow \tau\tau bbWW$   in the context of Type II 2HDM. The checkmarks indicate kinematically allowed channels. Also shown are the typical favored region of $\sin(\beta-\alpha)$  for each case (see Ref.~\cite{Coleppa:2013dya}).  }  
\label{tab:classification}
\end{table}

\begin{figure}[h!]
\centering
	\minigraph{7.cm}{-0.in}{(a)}{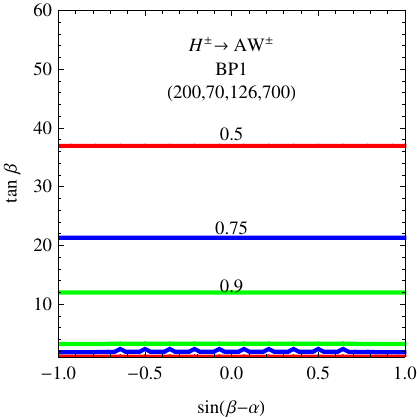}
	\minigraph{7.cm}{-0.in}{(b)}{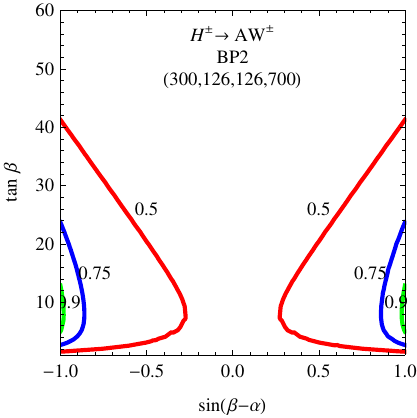}
	\minigraph{7.cm}{-0.in}{(c)}{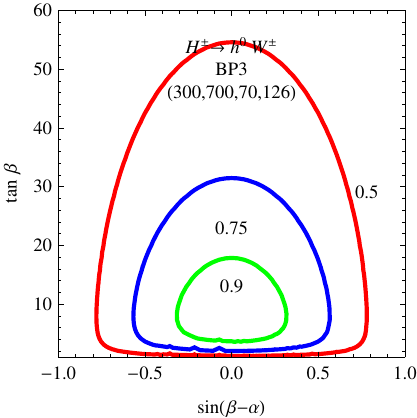}
	\minigraph{7.cm}{-0.in}{(d)}{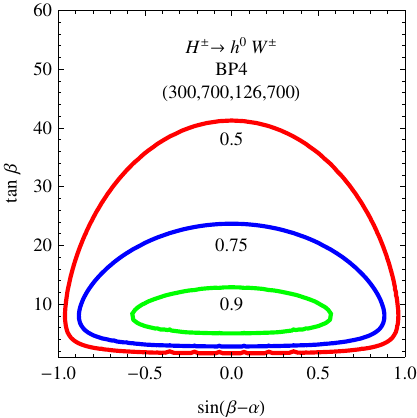}
\caption{Contours of branching fractions  of $\hc\to A\w$  [(a) and (b)] and $\hc\to \h\w$ [(c) and (d)] for each benchmark point.    
 }
\label{fig:BR_BP}
\end{figure}

In Fig.~\ref{fig:BR_BP}, we display the branching fraction of the $\hc\rightarrow A\w$ and $\h\w$ for the various benchmark points listed in Table~\ref{tab:classification} in the $\sba-\tan\beta$ plane.   For BP1 with $(m_{\hc},m_{\A}, m_{\h}, m_{\H})$=$(200, 70,126,700)$ GeV in panel (a),   BR$(\hc \rightarrow \A \w)$ is  independent of $\sba$, while decreasing at both large and very small $\tan\beta$, due to the competition of $\hc \rightarrow tb$ mode. 
BR($\hc\to\A\w$) can reach 90\% or larger in the range 3 $\lesssim\tan\beta\lesssim$  12.   Even for  $\tan\beta=$  37,  BR$(\hc \rightarrow \A \w)$ can be around  50\%. 

For BP2 with $(m_{\hc},m_{\A}, m_{\h}, m_{\H})$=$(300,126,126,700)$ GeV in panel (b),  BR$(\hc \rightarrow \A \w)$ decreases at small $|\sba|$ due to the opening of the $\hc\to\h\w$ channel.   BR$(\hc \rightarrow \A \w)$ is maximized for $\sba=\pm 1$ and intermediate $\tan\beta$, which is also the preferred region in the $\h$-126 case. 
 
For  BP3 with $(m_{\hc},m_{\A}, m_{\h}, m_{\H})$=$(300, 700,  70, 126)$ GeV in panel (c), maximal branching fraction for $\hc \rightarrow \h \w$ is obtained around $\sba=$ 0 where the coupling is maximal.   The decreasing of the branching fraction at large and small $\tan\beta$ is caused  by the enhanced $tb$ and $\tau\nu$ modes, while the decreasing of the branching fraction at $\sba \sim \pm 1$ is caused by the suppressed $\hc \rightarrow \h \w$ decay width as well as the enhanced $\hc \rightarrow \H W$ mode.  
 
For BP4 with $(m_{\hc},m_{\A}, m_{\h}, m_{\H})$=$(300, 700, 126, 700)$ GeV  in panel (d),  BR$(\hc \rightarrow \h \w)$ is suppressed at large $\tan\beta$ compared to BP3, since $\hc \rightarrow \h \w$ has more phase space suppression.    The reduction  of BR$(\hc \rightarrow \h \w)$  at larger $|\sba|$, however, is milder since $\hc \rightarrow \H \w$  is kinematically forbidden.   In the preferred regions   $\sba \sim \pm 1$ and $0.55 < \sba < 0.9$ (for  $1.5 < \tan\beta<4$) in the  $\h$-126 case,  BR$(\hc \rightarrow \h \w)$ is still large enough   to allow sensitivity in this channel. 

\subsection{Reach in parameter spaces}
To  translate the discovery and exclusion limits on $\sigma\times$BR in the $\tan\beta$ versus $\sba$ plane, we focus on the model implication for the $\tau\tau$ channel only since the limits for the $bb$ channel are too weak to be realized within the Type II 2HDM.

\begin{figure}[h!]
\centering
          \minigraph{7.cm}{-0.in}{(a)}{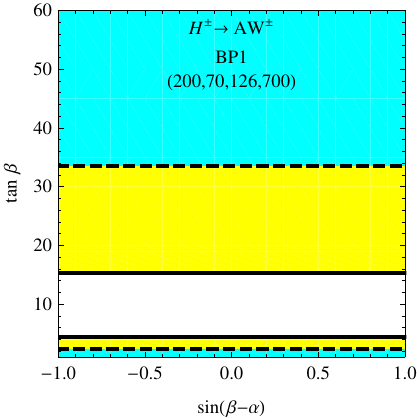}
          \minigraph{7.cm}{-0.in}{(b)}{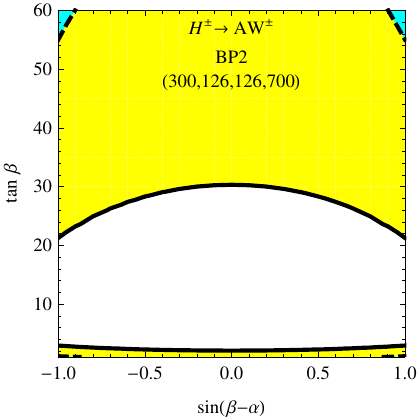}
         \minigraph{7.cm}{-0.in}{(c)}{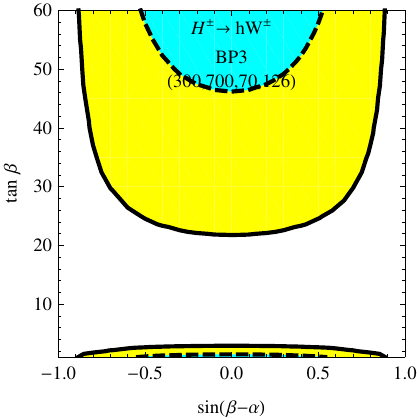}
          \minigraph{7.cm}{-0.in}{(d)}{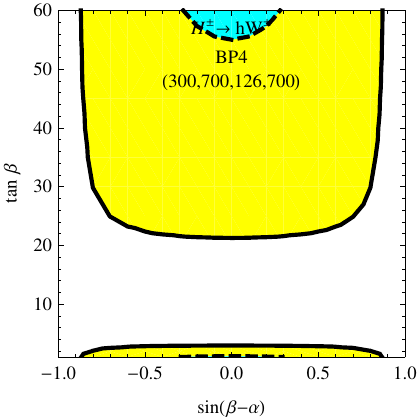}
	\caption{ The 95\% exclusion (yellow regions enclosed by the solid lines) and the $5\sigma$ discovery reach (cyan regions enclosed by the dashed lines) for $pp \rightarrow \hc tb \rightarrow \A\w tb/H\w tb\rightarrow   \tau\tau bbWW$  in the $\tan\beta$ versus $\sba$  plane for each benchmark point, with an integrated luminosity of 300 fb$^{-1}$ at the 14 TeV LHC.   }
\label{fig:tt-limit}
\end{figure}

In Fig.~\ref{fig:tt-limit}, we display the 95\% exclusion (yellow regions enclosed by the solid lines)  and 5$\sigma$ discovery limits (cyan regions enclosed by the dashed lines) for the various benchmark points at the 14 TeV LHC with 300 ${\rm fb}^{-1}$ integrated luminosity.  For BP1 with $\hc \rightarrow \A\w$ [panel (a)] ,   discovery is possible  for small $\tan\beta\lesssim$ 1.5 independent of $\sba$, and for large $\tan\beta\geq$  34.  The exclusion regions are much larger: $\tan\beta\lesssim$ 4 and $\tan\beta\gtrsim$  15.   Note that while the branching fraction is  relatively suppressed at small and large $\tan\beta$, as shown in Fig.~\ref{fig:BR_BP},   the $\hc$ production cross section is enhanced  in those regions, which is more than sufficient to offset the slightly reduced branching fractions.    Therefore,  we typically find exclusion and discovery regions appear in both the small and large $\tan\beta$ regions, so long as $\sigma\times$ BR values are large enough for exclusion/discovery.

The reach for BP2 [panel (b)] is  smaller compared to BP1 because of smaller cross sections associated with a 300 GeV $\hc$.  The model could still be excluded in quite a large range: $\tan\beta\lesssim$ 3, and $\tan\beta\gtrsim$ 22. These values, however, are dependent on $\sba$.    The maximum reach is achieved  around $\sba=\pm$ 1 where BR($\hc\to \A\w$) is maximized.  5$\sigma$ discovery, however, is not possible for this benchmark point except for very high $\tan\beta\geq$ 55, and $\sba\approx\pm$1.

For BP3 in panel (c), the reach is best for $\sba =$ 0: $\tan\beta\gtrsim  20$ or $\lesssim 3$ for 95\% C.L. exclusion and $\tan\beta\gtrsim  46$ or $\lesssim 1$ for 5$\sigma$ discovery.  The reach gets significantly weaker when $\sba$ approaches $\pm 1$ with the regions $|\sba|>$  0.9 providing no reach.    Note that for BP3 with $m_{\H}=126$ GeV, $\sba\approx$ 0 is also the favored region given the SM-like Higgs consideration.

BP4 is an interesting case as this corresponds to the charged Higgs decaying to a SM-like Higgs $\h$.   The exclusion reach is almost the same as in BP3, while the discovery reach is relatively weaker due to the suppression of the branching fractions at large or small $\tan\beta$, as shown in Fig.~\ref{fig:BR_BP} (d).   There  are small regions of parameter space around $\sba=0$ and $\tan\beta\gtrsim$ 55 or $\tan\beta\leq$ 1 that   permit discovery.   These exclusion or discovery regions  do not lie in the preferred region $\sba\approx\pm$1 for the $\h-$126 case. Note, however that the exclusion region for  $\hc \rightarrow \h \w$ is indeed sensitive to part of the region that is consistent with the observed Higgs signal:   $0.55 < \sba < 0.9$ with small $\tan\beta$~\cite{Coleppa:2013dya}.

\begin{figure}[h!]
\centering
	 \includegraphics[scale=1.0]{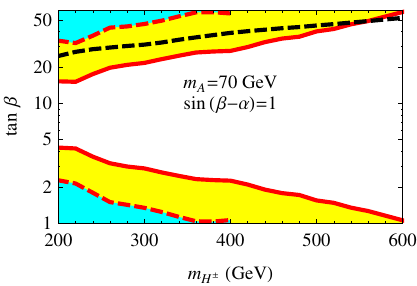}
	 \includegraphics[scale=1.0]{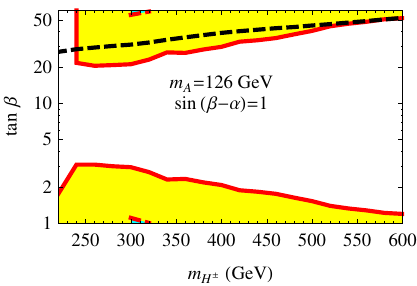}
	\caption{95\% exclusion (yellow regions bounded by solid red lines) and the $5\sigma$ discovery (cyan regions bounded by the dashed red lines) in the $m_{\hc}-\tan\beta$ parameter space for 300 fb$^{-1}$ luminosity in the $pp \rightarrow \hc tb \rightarrow \A\w tb \rightarrow   \tau\tau bbWW$  channel, with $m_\A=$  70 GeV (left panel) and 126 GeV (right panel).  Superimposed in black dashed line is the projected ATLAS  $\hc \rightarrow \tau\nu$ 5$\sigma$ discovery contours  with  100 fb$^{-1}$  luminosity.     $\sba$ is chosen to be 1 and $\H$ is decoupled.   }
\label{fig:mHtb}
\end{figure}

Fig.~\ref{fig:mHtb} shows the reach in the $m_{\hc}-\tan\beta$ for $\hc \rightarrow \A\w$, with $m_\A=$  70 GeV (left panel) and 126 GeV (right panel).   We have fixed $\sba=1$ and decoupled $\H$ such that  both $\hc\to\h\w, \H\w$ are absent. Superimposed on the plot in black dashed line is  the projected ATLAS $\hc \rightarrow \tau\nu$ discovery reach with 100 fb$^{-1}$ luminosity~\cite{ATLAS-PUB} for comparison.  The $m_{A}=$  70 GeV represents the best case scenario for discovery/exclusion.    While the reach in the exotic channel  $\hc\to \A\w$ is smaller compared  to the standard $\hc \rightarrow \tau \nu$ searches in the high $\tan\beta$ region,  $\A\w$ channel provides a reach in the small $\tan\beta$ regions  which is absent in the $\tau\nu$ mode.   Additionally, the model can be excluded at the 95\% C.L. for masses extending all the way to 600 GeV for both small and large $\tan\beta$ in this channel.   The $m_{A}=$ 126 case has limited sensitivity for discovery (constrained to only small regions 300 GeV $< m_{\hc}<   $ 320 GeV), but does provide an exclusion range that is comparable to the $m_{A}=$ 70  GeV case.

We conclude this section with the following observations:
\begin{itemize}
\item The best case scenario are the decays $\hc\to\A\w$ for the $\h-$126 case and $\hc\to\h\w$ in the $\H-$126 case for small daughter Higgs masses. 
\item The potentially interesting scenario $\hc\to \h\w$  with $\h$ being SM-like has sensitivity for 95\% C.L. exclusion at small and large $\tan\beta$ for $\sba$ different from $\pm$1.   The sensitivity for discovery, however, is constrained mostly to high $\tan\beta$ regions.  
\item There is sizeable  reach in both small and large $\tan\beta$ for exclusion for $m_{A}=$  70 GeV and $\sba=$ 1, while discovery is also possible for small $\tan\beta$ as  seen in Fig.~\ref{fig:mHtb}.  Specifically, discovery of the charged Higgs is possible for $m_{\hc}$ up to  400 GeV in both the small and large $\tan\beta$ regions.  
\item The reach in this exotic channel  $\hc \rightarrow \A\w/H\w$ is complementary to the conventional search channel  $\hc \to \tau\nu$, in particular, for small $\tan\beta$.   
\end{itemize}

\section{Conclusion}
\label{sec:conclusions}
 
The discovery of the Higgs at 126 GeV has not only confirmed the predictions of the SM, but has also ushered in a new era of discovery of beyond the SM physics.    Many  such scenarios incorporate an extended Higgs sector, which predict the existence of extra Higgs bosons other than the SM-like one.   Most of the current  searches for those extra Higgs bosons   focus on the conventional channels of $bb$, $\tau\tau$, $\gamma\gamma$, $WW$ and $ZZ$ for the neutral ones, and $\tau\nu$, $cs$ for the charged ones.  However, there have been efforts recently to study the exotic decay of these Higgs bosons to enhance their collider reaches~\cite{Coleppa:2014hxa, Brownson:2013lka,Maitra:2014qea,Basso:2012st,Dermisek:2013cxa, Curtin:2013fra,Tong_Hpm}.

Charged Higgses, compared to their neutral counterparts, are harder to discover.   This is mostly due to the relatively small associated production cross section of $\hc tb$ (compared to the gluon fusion process for the neutral ones), as well as the large SM backgrounds for the dominant decay mode  $\hc \to tb$.   The conventional search channel $\hc \to \tau \nu$ suffers from relatively small decay branching fraction and thus, it behooves us to consider other possible decays of the $\hc$ to enhance its reach at colliders.   In this paper, we analyzed the feasibility of discovering a charged Higgs boson in the process $\hc\to\A\w/H\w$, with the daughter Higgs decaying to either $\tau\tau$ or $bb$.

We obtained model independent limits on $\sigma\times{\rm BR}(pp\rightarrow \hc tb \rightarrow A\w  tb/H\w tb \rightarrow \tau\tau bbWW, bbbbWW)$ at the 14 TeV LHC.   For the $\tau\tau$ channel, we considered all three cases:  $\tau_{had}\tau_{had}$,  $\tau_{lep}\tau_{had}$, and  $\tau_{lep}\tau_{lep}$.   It turns out that  $\tau_{lep}\tau_{had}$ affords the best possible reach as we can take advantage of the same sign dilepton signal.  Combining all three channels, we find  for a daughter particle mass of  70 GeV, that the 95\% C.L. exclusion reach ranges from about  60 fb  to 25 fb, when $m_{\hc}$ is varied in the range 150 GeV$-$500 GeV with 300 ${\rm fb}^{-1}$ integrated luminosity at the 14 TeV LHC.  The 5$\sigma$ reach is about a factor of 3$-$4 higher.  This channel is statistically limited and the reach enhances with   increased luminosity.  The reach in the $bb$ channel is significantly worse.

We studied the implication of the $\sigma\times{\rm BR}$ reach  in the Type II 2HDM,  focusing on  $\hc\to\A\w$ and $\hc\to\h\w$ decays. We find that in this model, the $pp \rightarrow \hc tb   \rightarrow bbbbWW$ cross section  is too low for $\hc$ to be either discovered or excluded. However,  for the $\tau\tau$ mode, large regions of parameter space in $\tan\beta$ versus $\sin(\beta-\alpha)$ can be covered when the daughter Higgs mass is relatively light, in particular, for small and large $\tan\beta$.    The  exclusion region in the $m_{\hc}-\tan\beta$ plane can be extended to $m_{\hc}=$ 600 GeV, while discovery is possible for $m_{\hc}\lesssim $  400 GeV.  While the model can be excluded for a wide range of $\tan\beta$ values, discovery regions are mostly restricted to either small ($\lesssim$ 2) or large ($\gtrsim$  34 ) values.  Since the conventional search channel  $\hc \rightarrow \tau\nu$ is only sensitive to the large $\tan\beta$ region,  the exotic decay mode  $\hc \rightarrow A\w/H\w$ offers a complementary channel for charged Higgs searches.

Given the difficulties of the charged Higgs detection at hadron colliders, other search channels, for example, $qq^\prime \rightarrow \hc$, electroweak pair production of $H^+H^-$, $H^+W^-$, as well as charged Higgs produced in the decay of a heavy Higgs \cite{Hashemi:2013raa, THEO_WH_TB, Maitra:2014qea,Basso:2012st,Dermisek:2013cxa, Tong_Hpm,Christensen:2012si} should be studied to fully explore the discovery potential of the charged Higgses at the LHC.  A future lepton machine with high center of mass energy would certainly be useful for charged Higgs discovery.

\acknowledgments
We thank Peter Loch, John Paul Chou, John Stupak and Martin Flechl  for helpful discussions.  We also wishes to acknowledge the hospitality of the Aspen Center for Physics where part of the work was finished.  This work was supported by the Department of Energy under  Grant~DE-FG02-13ER41976.
 
  \bibliographystyle{JHEP}

\end{document}